\documentstyle[multicol,prl,aps,epsf,epsfig]{revtex}

\begin{document}

\title{Quantum Superarrivals}

\author{Somshubhro Bandyopadhyay$^1$, A. S. Majumdar\footnote{E-mail: archan@boson.bose.res.in}$^2$ and
Dipankar Home\footnote{E-mail: dhom@boseinst.ernet.in}$^1$}

\address{$^1$Department of Physics, Bose Institute 93/1 A. P. C Road, Calcutta
700009, India} 

\address{$^2$S. N. Bose National Centre for Basic Sciences,
Block JD,
Sector III, Salt Lake, Calcutta 700098, India}

\maketitle

\begin{abstract}

A curious effect is uncovered by calculating the  {\it time evolving} 
probability of reflection of a Gaussian wave packet  from a rectangular 
potential barrier {\it while} it is perturbed by  {\it reducing}  its height. 
A time interval is found during which this probability of reflection is 
{\it larger}  (``superarrivals'') than in the unperturbed case. 
This {\it nonclassical} effect can be explained by requiring a wave function  
to act as a ``field'' through which an action, induced by the perturbation 
of the  boundary condition,  propagates at a speed depending upon the rate 
of reducing the barrier height.

PACS number(s): 03.65.Bz

\end{abstract}

\begin{multicols}{2}

In recent years a number of interesting investigations have been reported on
the wave-packet dynamics\cite{greenberger}. Here we study it 
from a hitherto unexplored perspective. The reflection/transmission
probabilities for the scattering of wave packets by various obstacles are usually
considered from static or unperturbed potential barriers. Generally,
the time-independent
(asymptotic) values attained {\it after} a complete time evolution are calculated.
Here an interesting effect is   pointed  out  that occurs {\it during} the
time evolution. For this we consider the
dynamics of a wave packet scattered from a barrier {\it while} its
height is reduced to zero   {\it before} the asymptotic value of reflection
probability is reached.

For an unperturbed barrier, the reflection probability for an initially localized
wave packet  $\psi \left( x,t=0\right)$   is calculated by considering the
wave packet as a superposition of plane waves and by writing

\begin{equation}
\label{1}
\left| R_{0}\right| ^{2}=\int \left| \phi \left( p\right) \right| ^{2}\left| R\left( p\right) \right| ^{2}dp
\end{equation}
where  $\left| R\left( p\right) \right|^{2}$  is the reflection probability
corresponding to the plane wave component  $exp(ipx)$  and  $\phi (p)$
is the Fourier transform of an initial wave packet  $\psi (x,t=0)$. Since a
wave packet evolves in time,  $\left| R_{0}\right| ^{2}$  defined by Eq.
(1) denotes the {\it time-independent} value of reflection probability pertaining
to a wave packet, which is attained in the asymptotic limit ($t_{\infty}$)
of the time evolution. Thus $|R_{0}|^{2}$  can be expressed
in the following form

\begin{equation}
\label{2}
\left| R_{0}\right| ^{2}=\int ^{x\prime }_{-\infty }\left| \psi \left( x,t_{\infty }\right) \right|^{2}dx
\end{equation}
where $\psi \left( x,t_{\infty }\right)$   is an asymtotic form of the wave
packet attained by evolving from $\psi (x,t=0)$  and by being scattered
from a rectangular potential barrier of finite height and width. Note that $x\prime$  
lies at the left  of the initial profile of the wave packet such
that $\int ^{x\prime }_{-\infty }\left| \psi \left( x,t=0\right) \right|^{2}dx$
is negligible. Equivalence between the expressions (1) and (2) in the limit of large $t$ (compared to the time taken by the packet to get reflected from
the barrier) has been checked numerically. At any instant {\it before} the constant value
$\left| R_{0}\right|^{2}$ is attained, the time evolving reflection probability
in the region $-\infty <x\leq x\prime$   is thus given by

\begin{equation}
\label{3}
\left| R(t)\right|^{2}=\int ^{x\prime }_{-\infty }\left| \psi \left( x,t\right) \right|^{2}dx
\end{equation}

Now, suppose that {\it during} the time evolution of this wave packet, the barrier is
perturbed by reducing its height to zero within a very short
 interval of time
that is small compared to the time taken by the reflection probability to
attain its asymptotic value  $\left| R_{0}\right| ^{2}$. We compute the effects
of this ``sudden'' perturbation on $\left| R(t)\right|^{2}$.
The salient features are as follows:
(a) A finite time interval is found during which  $\left| R(t)\right|^{2}$
shows an {\it enhancement} ( ``superarrivals'')
in the perturbed case even though the barrier height is reduced. This time interval
and the amount of enhancement depend on the {\it rate} at which the barrier height
is made zero. 
(b) We show that the phenomenon of superarrivals is inherently quantum mechanical
by demonstrating that superarrivals {\it disappear} in the classical 
treatment of the problem.
(c) The {\it origin} of superarrivals may be understood by considering the wave
function to act as a ``field'' through which a disturbance 
from the ``kick'' provided by perturbing the barrier travels with a definite speed.    

In order to demonstrate the above features we begin by writing the initial
wave packet (in the units of $\hbar =1$  and $m=1/2$) 
\begin{equation}
\label{4}
\psi \left( x,t=0\right) =\frac{1}{\left[ 2\pi \left( \sigma _{0}\right) ^{2}\right] ^{1/4}}exp\left[ -\frac{\left( x-x_{0}\right) ^{2}}{4\sigma ^{2}_{0}}+ip_{0}x\right] 
\end{equation}
which describes a packet of width $\sigma _{0}$  centered around $x=x_{0}$ 
with its peak moving with a group velocity 
$v_g = 2p_{0}=\frac{\left\langle p\right\rangle }{m}$
towards a rectangular potential barrier. The point $x_{0}$  is chosen such
that $\psi \left( x,t=0\right)$ has a negligible overlap with the barrier.
For  computing $\left| R(t)\right|^{2}$ given by Eq. (3)
the time dependent Schrodinger equation is solved by using the numerical methods
developed by Goldberg, Schey and Schwartz\cite{goldberg}. In such a  treatment the
parameters are chosen in order to ensure that the spreading of a
packet is  negligible.
Here we choose $x_{0}=-0.3$, $\sigma _{0}=0.05/\sqrt{2}$  and $p_{0}=50\pi$.
The barrier is centered around $x_{c}=0$ with a width taken to be
0.016. For such a width, height of
the barrier (V) before perturbation is chosen to be $V=2E$, where $E$ is the
expectation value of the energy of the wave packet given by 
 $p^{2}_{0}+\frac{1}{4}\sigma ^{-2}_{0}$. This choice
ensures that:
(1) The reflection probability is close to $1$ 
since we are interested only in the reflection probability.
(2) At the same time  $V$ is not  too large . This  ensures that the 
reduction of the barrier height is not  too fast.

$\left| R(t)\right|^{2}$  is computed according to Eq. (3) by
taking various values of $x^{\prime}$ satisfying the condition
$x^{\prime } \le =x_{0}-3\sigma _{0}/\sqrt{2}$.
The computed evolution of $\left| R(t)\right|^{2}$ corresponds to the
building up of reflected particles with time. More precisely, it means that
a detector located within the region $-\infty < x < x^{\prime }$  measures
$\left| R(t)\right|^{2}$ by registering the reflected particles arriving
in that region up to various instants. 
First, we compute $\left| R(t)\right|^{2}$ for the wave packet scattered
from a static barrier $V=2E$. The relevant curve is shown in Figure 1
which tends towards a time-independent value which is the stationary state
reflection probability $\left| R_{0}\right| ^{2}$ given by Eq.(1), or
equivalently by Eq.(2).  Next, 
we proceed to study the consequence of reducing the
barrier height from $V=2E$  to $V = 0$. The time evolution of $\left| R(t)\right|^{2}$ 
in this case is 
studied by varying the ways in which the barrier height is reduced.

In  the specific cases  studied, the potential $V$  goes to zero linearly within
a switching off time $\epsilon$  starting at time $t=t_{p}$  chosen
to be $8\times 10^{-4}$ (note that numbers denoting the various instants
are in terms of time steps; for example, $t=8\times 10^{-4}$ corresponds
to 400 time steps). Here $\epsilon \ll t_{0}$, $t_{0}$  being the time
required for $\left| R(t)\right|^{2}$  to attain the asymptotic value $\left| R_{0}\right|^{2}$.
The short time span $\epsilon$ over which the perturbation takes place
is given by $\left[ t_{p},t_{p}+\epsilon \right]$.
$t_p$ is chosen such  that at that instant the overlap of the wave packet with the
barrier is significant.
Figure 1 shows the evolution of $\left| R(t)\right|^{2}$ for various values
of $\epsilon$.  Varying $\epsilon$   signifies changing
the time span over which the barrier height goes to zero which in turn means
different rates of reduction.  Figure 1  reveals that 
\begin{equation}
\label{5}
\left| R_{p}(t)\right|^{2}=\left| R_{s}(t)\right| ^{2}\quad \quad \quad t\leq t_{d}
\end{equation}
\begin{equation}
\label{6}
\left| R_{p}(t)\right|^{2}>\left| R_{s}(t)\right| ^{2}\quad \quad \quad t_{d}<t\leq t_{c}
\end{equation}
\begin{equation}
\label{7}
\left| R_{p}(t)\right| ^{2}<\left| R_{s}(t)\right| ^{2}\quad \quad \quad t>t_{c}
\end{equation}
where 
$t_{c}$ is the instant when the two curves cross each other, and $t_{d}$ 
is the time  from which the curve corresponding to the perturbed case starts
deviating from that in the unperturbed case. Here $t_{c}>t_{d}>t_{p}$.
\begin{figure}
\begin{center}
\centerline{\epsfig{file=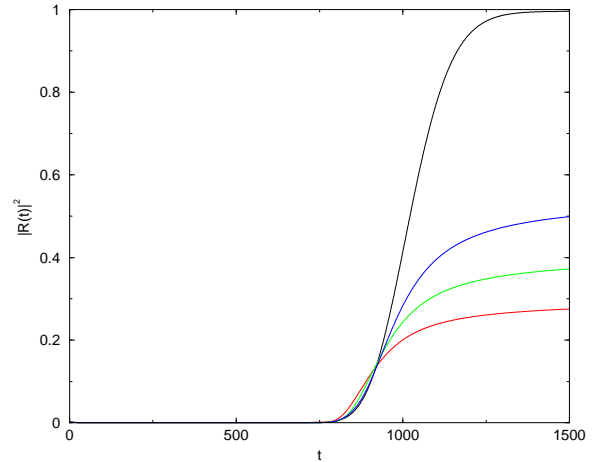,height=3.0in,angle=270}}
\narrowtext{\caption{ The top curve corresponds to the static case and 
reaches value $1$ asymptotically. $|R(t)|^2$ for other curves correspond
to various values of
$\epsilon$. The curve with the lowest
asymptotic value corresponds to the smallest value of $\epsilon$
chosen for this set. As one increases $\epsilon$, superarrivals are
slowly wiped off.}}
\end{center}
\end{figure}
As the
barrier height is made zero, one does not expect at any time an increase in
the reflected particle flux compared to that in the unperturbed case. Nevertheless,
the inequality (6) shows that there is a finite time interval 
$\Delta t \equiv t_c-t_d$  
during which the probability of finding  reflected particles is
{\it more}
(superarrivals) in the perturbed case than when the barrier is left unperturbed
(see Figure 1). A detector placed in the region $x<x'$ would therefore
register {\it more} counts during this time interval $\Delta t$ even
though the barrier height had been {\it reduced} to zero {\it prior} to that.
This effect is essentially quantum mechanical.
In order to explicitly show this, we consider an initial
distribution of particles given by  the spreads in both position and momentum
corresponding to a Gaussian wave packet. Such
particles are now assumed to obey the classical equation of motion. We solve
numerically the Liouville equation in this same time-varying situation in order to obtain 
the time-dependent number density of particles at the detector. Our results
are plotted in Figure 2 where we show that there are {\it no} superarrivals for
three different values of $\epsilon$.

 In order to have a {\it quantitative measure} of superarrivals
we define the parameter $\eta$  given by
\begin{equation}
\label{8}
\eta =\frac{I_{p}-I_{s}}{I_{s}}
\end{equation}
where the quantities $I_{p}$  and $I_{s}$  are defined with respect to
$\Delta t$ during which superarrivals occur
\begin{equation}
\label{9a}
I_{p}=\int _{\Delta t}\left| R_{p}(t)\right|^{2}dt
\end{equation}
\begin{equation}
\label{9b}
I_{s}=\int _{\Delta t}\left| R_{s}(t)\right|^{2}dt
\end{equation}
\begin{figure}
\begin{center}
\centerline{\epsfig{file=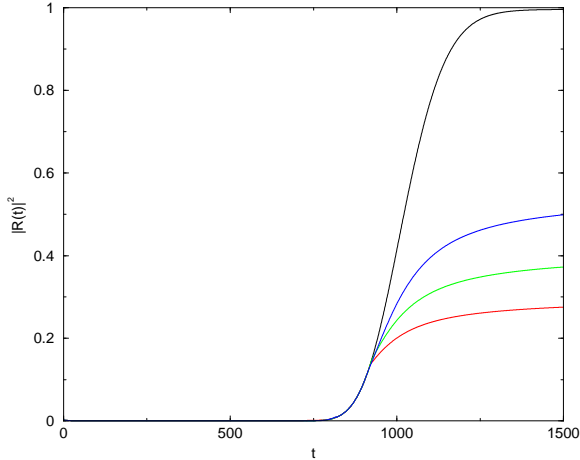,height=3.0in,angle=270}}
\narrowtext{\caption{The time-varying reflection probability for the classical evolution
is plotted for the same values of
$\epsilon$ as in Figure 1. The absence of superarrivals in this case
demonstrates the nonclassical nature of this phenomenon.}}
\end{center}
\end{figure}  

We plot the variation of $\eta$   
with respect to $\epsilon$  for three different detector
positions in Figure 3. The effect of reducing the barrier width on the
magnitude of superarrivals is shown in figure 4.
\begin{figure}
\begin{center}
\centerline{\epsfig{file=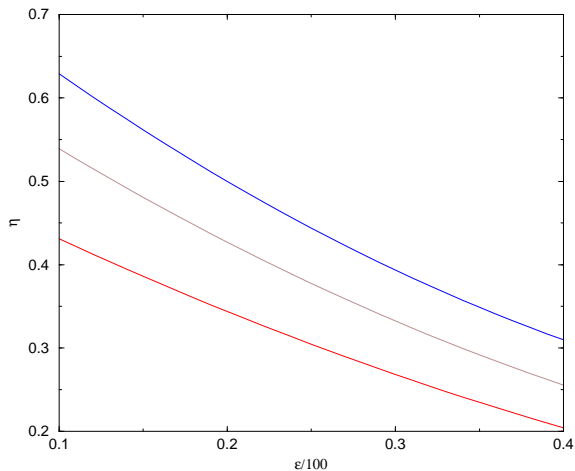,height=3.0in,angle=270}}
\narrowtext{\caption{The magnitude of superarrivals $\eta$ diminish
with an increase in $\epsilon$, the time taken for barrier height reduction.
This behaviour is seen for three different detector positions $x'$= -0.4,
-0.5 and -0.6 respectively.}}
\end{center}
\end{figure}  
\begin{figure}
\begin{center}
\centerline{\epsfig{file=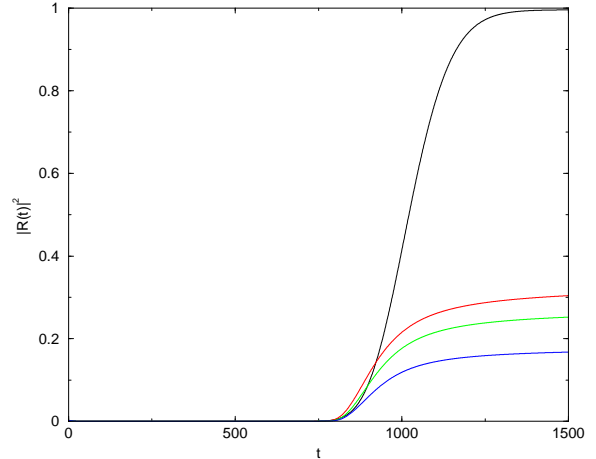,height=3.0in,angle=270}}
\narrowtext{\caption{Here we show that superarrivals diminish by decreasing
the width of the barrier. They  completely disappear for a small enough
barrier width. The three curves for the perturbed cases correspond to the widths
of 0.016, 0.008, and 0.004 respectively.}}
\end{center}
\end{figure}
The results obtained from Figures 1--4 can be summarized as follows: 
(a) There exists a finite time interval $\Delta t$  during which
an {\it increase} in the reflection probability (superarrivals) occurs
for the perturbed cases compared to the unperturbed situation. 
(b) Superarrivals are inherently {\it nonclassical}.
(c) The magnitude of superarrivals $\eta$
is appreciable only in cases where the wave packet has some {\it significant overlap} with
the barrier during its switching off. Both $\eta$ and $\Delta t$ (duration 
of superarrivals) {\it fall off} with {\it increasing} $\epsilon$.
(d) Superarrivals given by $\eta$ gradually reduce to {\it zero} upon {\it decreasing} the
barrier width, while keeping the initial barrier height $B$ fixed.
(e) Superarrivals {\it persist}, as we have checked, even
if the detector is placed at a distance $x'$ greater than $8\sigma$
to the left of the initial position of the centre of the wave
packet $x_0$.

Next, we consider the  question as to {\it how fast}
the influence of barrier perturbation travels across the wave packet (signal
velocity $v_e$). 
Note that even in a classical theory the
information content of a wave packet does {\it not} always propagate with
the group velocity $v_g$ of a wave packet which is usually identified
with the velocity of the peak of a wave packet\cite{sommerfeld}. 
 Profiles of the quantum wave packet are plotted at various instants
in Figure 5. An incident packet gets distorted upon hitting the
time-varying barrier. It splits into two pieces, one of which moves towards
the right (transmitted particles). The reflected packet has
 a secondary peak shifted towards the left. It is thus not possible
to uniquely define a group velocity $v_g$ for the reflected packet in this case.

 The action due to a local perturbation (reduction of barrier height) 
propagates across the wave packet with a signal velocity
 $v_{e}$  which affects the time evolving reflection
probability  that
can be measured at different points. Thus a distant observer who records the
growth of reflection probability becomes aware of the perturbation of the barrier
(occuring from an instant $t_{p}$) at the instant $t_{d}$  when
the time varying reflection probability starts deviating from that
in the unperturbed case. Then $v_{e}$  is given by 
\begin{equation}
v_{e}=\frac{D}{t_{d}-t_{p}}
\end{equation}
\begin{figure}
\begin{center}
\centerline{\epsfig{file=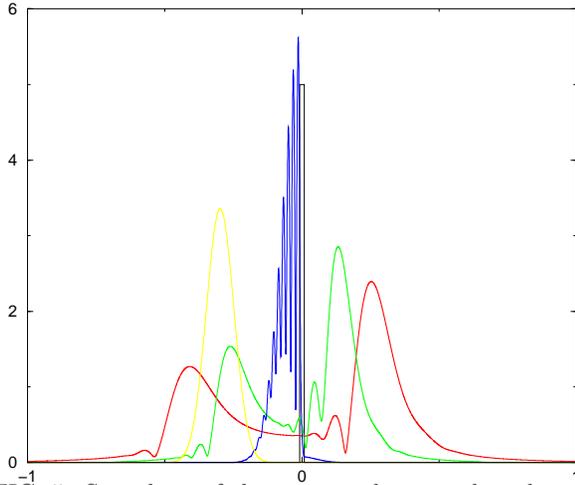,height=3.0in,angle=270}}
\narrowtext{\caption{Snapshots of the wave packet are plotted at four 
different instants of time. The initial narrow Gaussian is heavily distorted
upon striking the barrier. It splits up into two, with the reflected part
possessing a secondary peak shifted  towards the detector.}}
\end{center}
\end{figure}
We compute $v_e$ and $v_g \equiv <p>/m$ for a range of parameters and
plot $v_e/v_g$ versus $\epsilon$ in Figure 6.
Both $\Delta t$ (the duration of 
superarrivals) and $v_e$ (the signal velocity) {\it decrease} with 
{\it increasing} $\epsilon$
(or, decreasing rate of perturbation).
The magnitude of superarrivals ($\eta$) also {\it decreases} with {\it increasing} $\epsilon$ (Figure 3).
 From such behaviours of $\eta$, $\Delta t$
and $v_e$ we infer the following {\it explanation} for the origin of 
superarrivals. The barrier perturbation imparts a ``kick'' on the impinging 
wave packet which spits, and a part of it is reflected with a distortion.
A finite disturbance proportional to this ``kick'' or the rate of perturbation propagates
from the reducing barrier to the reflected
packet, which results in a proportional  magnitude of superarrivals $\eta$.
Note that information about the barrier perturbation  reaches the
detector at the instant $t_d$ with a velocity $v_e$ which  decreases with the
decreasing magnitude of impulse imparted to a wave packet.  These results 
therefore suggest
that  information about the barrier perturbation propagates with a 
{\it definite speed} across the
wave function which plays the role of a ``field''.
\begin{figure}
\begin{center}
\centerline{\epsfig{file=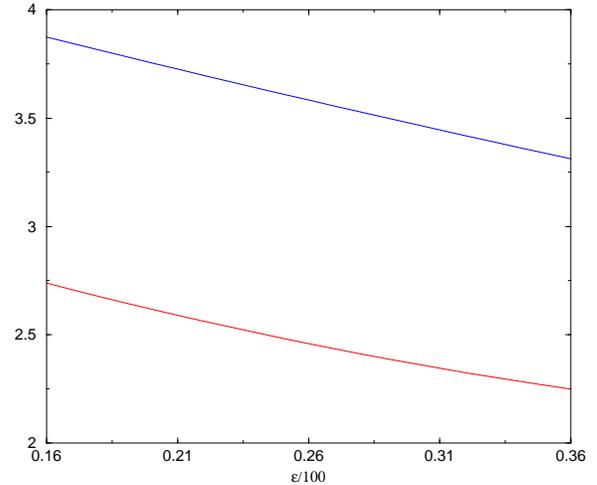,height=3.0in,angle=270}}
\narrowtext{\caption{The upper curve represents a plot of $\Delta t$ 
(duration of superarrivals) versus $\epsilon$. The lower curve is a plot of $v_e/v_g$ versus
$\epsilon$. Here the detector position $x' = -0.4$. }}
\end{center}
\end{figure}
To sum up, superarrivals stem essentially from the objective reality of
a wave function acting as a ``field'' which mediates the propagation of a
physical disturbance induced by the barrier perturbation. This disturbance
propagates with a  speed that varies according to the rate at which
the barrier is perturbed. Thus the  phenomenon of superarrivals has
a distinct quantum mechanical significance. Its ramifications call for further
studies. In particular,  different types of
perturbations  may be studied to probe  the
viability of single particle experiments\cite{electron,neutron} for
demonstrating this effect.

\vskip 0.1in

D. H. acknowledges helpful discussions with Shyamal Sengupta and 
Brian Pippard. The support provided by the Dept. of Science
and Technology, Govt. of India is also acknowledged.

\end{multicols}

\end{document}